\documentstyle[12pt]{article}
\title{Dispersion Relations and the Nucleon Polarizability$^*$}
\author{Barry R. Holstein\\
Department of Physics and Astronomy\\
University of Massachusetts\\
Amherst, MA  01003\\
\and
A. M. Nathan\\
Nuclear Physics Laboratory and Department of Physics\\
University of Illinois at Urbana-Champaign\\
Champaign, IL  61820}
\begin{document}
\begin{titlepage}
\maketitle
\begin{abstract}
Recent experimental results on the proton and neutron polarizabilities are
examined from the point
of view of backward dispersion relations.  Results are found to be in
reasonable
agreement with the measured values.  A rigorous relationship
between the nucleon and pion polarizabilities is derived and shown to be
in excellent agreement with several models.
\end{abstract}
\vfill
${}^*$Research supported in part by the National Science Foundation
\end{titlepage}
%\pacs{14.20.Dh, 13.40.-f}

\section{ Introduction}

There has been a good deal of recent experimental activity involved with
measurement of the electric and magnetic polarizabilities of the nucleon,
labeled $\bar{\alpha}_E$ and $\bar{\beta}_M$, respectively.  As a result
there
now exist reasonably precise values for both the proton \cite{Fe91,Zi92,%
Ha93} and the
neutron \cite{Sc91}
\begin{eqnarray}
\bar{\alpha}_E^p&=& (11.3 \pm 0.7 \pm 0.8)
\qquad
   \bar{\beta}_M^p= (2.9 \mp 0.7 \mp 0.8)
	\nonumber\\
\bar{\alpha}_E^n &=& (12.6 \pm 1.5 \pm 2.0)
\qquad
\bar{\beta}_M^n = (3.2 \mp 1.5 \mp 2.0).
\nonumber\\
\label{eq:polarizabilities}
\end{eqnarray}
Above and hereafter all polarizabilities are quoted in units of $10^{-4}$
fm$^3$.  For the proton, the first error is the combined statistical and
systematic uncertainty
based on combining the results of several experiments, and the second
represents an estimated theoretical error based on the model dependence in
the extraction of the polarizabilities from the Compton scattering cross
sections \cite{Na94}.  For the neutron, the first error is statistical and the
second is systematic.
While there remain nonnegligible experimental uncertainties, it appears
likely
that the neutron and
proton magnetic polarizabilities are nearly identical while the electric
polarizability of the neutron is
slightly larger than that of the proton.  The former result is not
unexpected.  However, the latter is
somewhat of a surprise, at least in the context of a simple nonrelativistic
valence quark model for the nucleon.  In such a model, the polarizabilities
are given by \cite{Fr75}
\begin{equation}
	\bar{\alpha}_E=\alpha_E+\Delta\alpha_E,\qquad
	\bar{\beta}_M=\beta_M+\Delta\beta_M,
\label{eq:bars}
\end{equation}
where
\begin{eqnarray}
\alpha_E&=&2\alpha\sum_{n\ne 0}{|\langle n|\sum_ie_i({\bf r}_i-
{\bf
R}_{CM})_z|0\rangle |^2
\over E_n-E_0}\nonumber\\
\beta_M&=&2\alpha\sum_{n\ne 0}{\langle n|\sum_i{e_i\over
2m_i}(\sigma_{iz}+L_{iz})|0\rangle |^2\over
E_n-E_0},
\label{eq:pol0}
\end{eqnarray}
and
\begin{eqnarray}
\Delta\alpha_E&=&{Q\alpha\over 3M}\langle\sum_ie_i({\bf r}_i-{\bf
R}_{CM})^2\rangle
\nonumber\\
\Delta\beta_M&=&-{\alpha\over 2M}\langle\left(\sum_ie_i({\bf r}_i-{\bf
R}_{CM})\right)^2\rangle -{\alpha\over 6}\langle\sum_i{e_i^2({\bf r}_i-{\bf
R}_{CM})^2\over
m_i}\rangle,
\label{eq:deltas}
\end{eqnarray}
where Q is 1 for the proton and 0 for the neutron.
In Eq.~(\ref{eq:pol0}) the sum rule component,
which is usually called the paramagnetic
polarizability, receives
its most important
contribution from the Delta
intermediate state, which is an isovector excitation and therefore
contributes equally to the neutron and
proton.  One finds, including only this contribution \cite{Mu93}
\begin{equation}
\beta_M^{\circ,p}=\beta_M^{\circ,n}\approx 13.
\label{eq:beta0}
\end{equation}
The term $\Delta\beta_M$, which is usually called the diamagnetic
polarizability, can be estimated within a simple
nonrelativistic valence constituent quark model, with Hamiltonian
\begin{equation}
H={1\over 2m}({\bf p}_1^2+{\bf p}_2^2+{\bf p}_3^2)+{m\omega_0^2\over 2}
({\bf r}_{12}^2+{\bf r}_{13}^2+{\bf r}_{23}^2),
\end{equation}
resulting in \cite{Ho92}
\begin{eqnarray}
\Delta\beta_M=-{\alpha\over m\omega_0}\left\{\begin{array}{ll}
                                                  {2\over 3} & p \\
                                                  {5\over 9} & n
                                             \end{array}\right.
=-\alpha \langle r_p^2\rangle\left\{\begin{array}{ll}
                       {2\over 3} & p \\
                       {5\over 9} & n
                \end{array}\right.\nonumber\\
{\rm where}\quad <r_p^2>={2\over 3}{1\over m\omega_0}\quad
\mbox{is the proton charge radius.}
\label{eq:deltabeta}
\end{eqnarray}
Such an approach is clearly unrealistic.  Indeed in such a picture
the nucleon and Delta
are degenerate and the neutron has zero charge radius.  However, if a
spin-spin interaction is
included these problems can be ameliorated but Eq.~(\ref{eq:deltabeta}) is
only
slightly
affected\cite{Ho92}:
\begin{equation}
\Delta\beta_M=-\alpha(\langle r_p^2\rangle -\langle r_n^2\rangle)\times
\left\{\begin{array}{ll}
                                                    {2\over 3} & p \\
                                                    {5\over 9} & n
                                              \end{array}\right.
\approx                \left\{\begin{array}{ll}
                              -10.2 & p \\
                              -8.5  & n
                             \end{array}\right.
\label{eq:delbeta}
\end{equation}
Thus we anticipate
\begin{equation}
\bar{\beta}_M^p-\bar{\beta}_M^n\approx 1.7.
\end{equation}
Combining Eq.~(\ref{eq:beta0}) and Eq.~(\ref{eq:delbeta}), we find
\begin{equation}
\bar{\beta_M}\approx
                             \left\{\begin{array}{ll}
                              2.8 & p \\
                              4.5 & n
                             \end{array}\right.
\label{eq:beta}
\end{equation}
in reasonable accord with experiment.
However, this
agreement should not be overemphasized, as the Delta is only the
most important of a large
number of possible intermediate states and the use of a simple valence
quark model is also open to question.

What {\it is} a problem is the electric polarizability in this simple model,
for which the recoil contribution is
given by \footnote{In a relativistic treatment, $\Delta\alpha_E$ has an
additional contribution ${\alpha(\lambda^2+Q)\over 4M^3}$, where
$\lambda$ is the anomalous magnetic moment \cite{Lv93a}.  Numerically,
this extra term is 0.71 and 0.62 for the proton and neutron, respectively.}
\begin{equation}
\Delta\alpha_E={\alpha\over 3M}\left\{\begin{array}{lll}
                                       \langle r_p^2\rangle  & p \\
                                       0  & n
                                     \end{array}\right.
=                            \left\{\begin{array}{ll}
                                 3.6 & p \\
                                 0 & n
                               \end{array}\right.
\end{equation}
while the sum rule in Eq.~(\ref{eq:pol0}) gives\cite{Pe81}
\begin{equation}
\alpha_E^{p}=\alpha_E^{n}={2\alpha\over
3\bar{\omega}}\langle r_p^2\rangle  \approx
10.8,
\end{equation}
where we have used a closure approximation and an average nucleon
excitation energy of $\bar{\omega}\approx$ 600 MeV.  The equality between
the neutron and proton values is a result of charge symmetry, which requires
that
the valence quark excitations lead to identical excited states and p,n matrix
elements.  The precise value of
the sum, which is difficult to calculate,
then cancels out when we take the difference
\begin{equation}
\bar{\alpha}_E^n-\bar{\alpha}_E^p\approx\Delta\alpha_E^n-
\Delta\alpha_E^p
=-3.6.
\label{eq:quark}
\end{equation}
This expectation, however, is in strong opposition to the experimental
indication that
\begin{equation}
\bar{\alpha}_E^n-\bar{\alpha}_E^p\ge 0.
\end{equation}

In fact, this difficulty is just another example of a well-known problem with
valence quark models for the structure of the nucleon:  namely, chiral
symmetry is badly
broken in such models because of the omission of mesonic degrees of freedom.
By including only valence quark excitations in the sum over
intermediate states, we are forced into the conclusion that the electric
polarizability of the proton exceeds that of the neutron, in direct
disagreement
with experiment.  In fact, it is well known that in the threshold region, the
pion photoproduction on the nucleon is primarily nonresonant and that the
cross section on the neutron is about 30\% larger than that on the proton, a
result which can easily be understood from a consideration of the effective
dipole moment of the $\pi$N system \cite{We88}.  Similar considerations
lead us to expect that $\alpha_E$ for the neutron will exceed that
for the proton.  This qualitative idea is supported in part by a calculation in
the context of the cloudy bag model \cite{We85}, where it was shown that
both the
electric and the diamagnetic polarizabilities are dominated by the
polarization of the
pion cloud relative to the quark core and have very little contribution from
the
polarization of the core itself.  This would lead one to expect the neutron
electric polarizability to exceed that of the proton, although a definitive
quantitative
calculation in that model is not possible.  A reasonable estimate
{\it is} possible in chiral
perturbation theory, where the only degrees are freedom that matter are
pionic, and a recent one-loop calculation yields results in reasonable
agreement
with
experiment\cite{Be91}
\begin{eqnarray}
(\alpha_E^n-\alpha_E^p)^{\chi pt}&=&3.1\nonumber\\
(\beta_M^n-\beta_M^p)^{\chi pt}&=&0.3.
\label{eq:chiral}
\end{eqnarray}
However, here too a rigorous evaluation is not available, as inclusion of
important contributions
such as the Delta are two-loop in character and are outside the present
calculational framework \cite{Lv93}.

We conclude then that a simple valence quark picture of the nucleon
is in disagreement with experiment and that
inclusion of meson cloud effects is {\it required} in order to understand
the result that $\bar{\alpha}_E^n>\bar{\alpha}_E^p$.  This
finding is similar to that in the interpretation of the
$\langle N | \bar{s} \gamma_\mu\gamma_5 s | N \rangle$ matrix
element, which also
vanishes in a valence quark model but can be understood qualitatively by
inclusion of a kaon cloud
via $N\rightarrow \Lambda K\rightarrow N$ \cite{Ho90}.  However, a
reliable
calculation of the polarizability and of the strangeness
content in this fashion is not possible.

Instead we follow a completely different approach, that of dispersion
relations.  On the one hand, this technique is capable of complete rigor in
that
the relations depend only on unitarity and certain analytic properties of the
Compton scattering amplitudes.  On the other hand, it is
semi-phenomenological in that the evaluation of the dispersion integrals
requires
as input either experimental data or some reasonable theoretical {\it ansatz}
when
the required data are not available.  In the next section we will present our
results on
the evaluation of
the so-called backward dispersion relation for $\bar{\alpha}_E-
\bar{\beta}_M$.  Then we show how this dispersion relation can be used
to calculate the contribution of the polarizability of the pion to that of
the nucleon.  Our conclusions are summarized in the concluding section.

\section{ Dispersion Sum Rules for the Polarizabilities}

By combining dispersion relations with low energy theorems  for the Compton
scattering amplitudes, one can derive sum rules for the polarizabilities.  A
comprehensive review of this subject has been given by Petrun'kin
\cite{Pe81}.  The best known sum rule, the so-called Baldin-Lapidus sum
rule,  is based on the forward dispersion relation for the spin-independent
part of the Compton scattering amplitude \cite{Ba60}:
\begin{equation}
\bar{\alpha}_E+\bar{\beta}_M={1\over 2\pi^2}\int_0^\infty
{d\omega\sigma_{\rm tot}(\omega)\over {\omega}^2}
=   \left\{\begin{array}{ll}
         (14.2\pm 0.5)&\mbox{ [proton]} \\
         (15.8 \pm 0.5)&\mbox{ [neutron]},
       \end{array}\right.
\label{eq:sum}
\end{equation}

In this expression, $\sigma_{\rm tot}$ is the total photoabsorption cross
section, and the numerical values are based on the tabulations of those cross
sections for the proton and neutron \cite{Da70,Lv79}.  The numbers given in
Eq.~(\ref{eq:sum}) were
actually used as a constraint in obtaining the experimental results in
Eq.~(\ref{eq:polarizabilities}), so that those results do not test this sum
rule.
However, it {\it is} possible to reanalyze the recent data for the proton
without
imposing the sum rule constraint, in which case one obtains \cite{Na94}
\begin{eqnarray}
\bar{\alpha}_E^p+\bar{\beta}_M^p&=&12.0\pm 2.3 \mbox{ [experiment],}
\end{eqnarray}
verifying the sum rule at the 1-standard deviation level.

It is also possible to write down sum rules for the difference of the electric
and magnetic polarizabilities.  The one we consider here is the so-called
backward sum rule, which is based on
a 180$^\circ$ dispersion relation and has the form\cite{Be74}:
\begin{equation}
\bar{\alpha}_E-\bar{\beta}_M \, = \, \mbox{(s-channel contribution)}\,+\,
\mbox{(t-channel contribution)}.
\label{eq:diff}
\end{equation}
The s-channel contribution is similar to Eq.~(\ref{eq:sum}), with a
relativistic
correction and with contributions from excitations
with opposite parity entering with opposite
sign---
\begin{eqnarray}
\mbox{s-channel contribution}={1\over 2\pi^2}\int_0^\infty
{d\omega\over \omega^2}(1+{2\omega\over m})^{1\over
2}\left[\sigma_{\rm tot}\mbox{($\Delta$P={\sc yes})}-\sigma_{\rm
tot}\mbox{($\Delta$P={\sc
no})}\right],\nonumber\\
\quad
\label{eq:schan}
\end{eqnarray}
where $\sigma_{\rm tot}$($\Delta$P={\sc yes}) and $\sigma_{\rm
tot}$($\Delta$ P={\sc
no})
represent those
pieces of  $\sigma_{\rm tot}$ arising from
multipoles which change and do not change parity, respectively.  The
t-channel contribution can be written as
\begin{eqnarray}
\mbox{t-channel contribution}={1\over 64\pi^2}\int_{4m_\pi^2}^\infty
{dt\over t^2}\sqrt{t-4m_\pi^2\over t}\int d\Omega
\left[A^{(+)}(t,\cos\theta )
\right.\nonumber\\
\qquad+\left.im\sqrt{t-4m_\pi^2\over 4m^2-t}\cos\theta
B^{(+)}(t,\cos\theta )\right]
F^*_0(t,\cos\theta ),
\label{eq:tchan}
\end{eqnarray}
which corresponds to the approximation of including only the $N\bar{N}
\rightarrow\pi\pi\rightarrow\gamma\gamma$ intermediate state.  The
integration variable t is the square of the total center-of-mass energy of the
$\gamma\gamma$ system.
Here $A^{(+)}, B^{(+)}$ are the conventional CGLN isospin-even amplitudes
for $N\bar{N}\rightarrow\pi\pi$ \cite{Ch57}, $F_0(t, \cos\theta )$ is the
I=0 Gourdin-Martin $\gamma\gamma\rightarrow\pi\pi$ amplitude
\cite{Go60}, and m and m$_\pi$ are the nucleon and pion masses,
respectively.
Because of the restriction to isoscalar amplitudes required by
G-parity invariance,
only even partial waves are permitted.  Including only S- and D-waves,
Eq. \ref{eq:tchan} simplifies to the form
\begin{eqnarray}
\mbox{t-channel contribution}={1\over 16\pi^2}\int_{4m_\pi^2}^\infty
{dt\over t^2}{16\over 4m^2-t}\sqrt{t-4m_\pi^2\over t}\left[
f_+^0(t)F_0^{0*}(t)
\right.\nonumber\\
\qquad-\left.(m^2-{t\over 4})({t\over 4}-m_\pi^2)f_+^2(t)F_0^{2*}(t)\right],
\label{eq:tchan1}
\end{eqnarray}
where the partial wave helicity amplitudes $f_+^J(t)$ for
$N\bar{N}\rightarrow\pi\pi$ are given by Frazer and
Fulco \cite{Fr60} while the corresponding partial wave amplitudes
$F_0^J(t)$ for $\gamma\gamma\rightarrow\pi\pi$ are defined in
Ref.~\cite{Go60}.

The backward sum rule has been previously evaluated by several authors
\cite{Be74,Gu78,Bu79}.  However, several recent developments have renewed
interest in this
sum rule and have motivated us to perform a reanalysis
with an eye towards a meaningful
confrontation with the new experimental values.  Such a
confrontation is now possible because, as we will point out shortly,
developments in understanding of the
$\gamma\gamma\rightarrow\pi\pi$ process has removed a major
uncertainty in the calculation.  Also the recent recognition of the
importance of pions has rekindled interest in the relationship between the
nucleon and pion polarizabilities \cite{Bu79}, an issue that we will
specifically address later in this paper.  We now describe in detail our
calculation.

In principle the s-channel integral is straightforward to calculate---provided
one knows the
multipole decomposition of $\sigma_{\rm tot}$, one can
separate the $\Delta$P={\sc yes} from the $\Delta$P={\sc no}
contributions.  Such a decomposition has been
performed, however,  only for the
$\pi N$ final state, which we therefore treat separately from the multi-pion
final states.  For this $\pi N$ final state, which dominates
$\sigma_{\rm tot}$ below 500 MeV, we use the multipoles of the VPI\&SU
group \cite{Ar90} and include $\pi N$ partial waves through L=4.  The
integrands for the proton are
shown in Fig.~\ref{fig:schan}, and the corresponding integrands for the neutron
are
quite similar.  Integrating up to 1800~MeV, we obtain the results
\begin{eqnarray}
\bar{\alpha}_E^p-\bar{\beta}_M^p&=&\left\{\begin{array}{ll}
                                 +4.80 &  \mbox{[s-channel single pion-
$\Delta$P={\sc yes}]} \\
                                -10.78 &  \mbox{[s-channel single pion-
$\Delta$P={\sc no}]}
                              \end{array}\right.
\nonumber\\
\bar{\alpha}_E^n-\bar{\beta}_M^n&=&\left\{\begin{array}{ll}
                                 +6.04  & \mbox{[s-channel single pion-
$\Delta$P={\sc yes}]} \\
                                 -11.31 & \mbox{[s-channel single pion-
$\Delta$P={\sc no}].}
                                 \end{array}\right.
\label{eq:onepi}
\end{eqnarray}

For the multi-pion contribution, a precise calculation is not possible since
an experimental multiple decomposition has not yet been performed.
Nevertheless, one can establish rigorous bounds
on that contribution in the following manner.
At any given
energy the entire multi-pion contribution to $\sigma_{\rm tot}$ can be
determined by subtracting the
{\it calculated} value of the single-pion contribution (using the VPI\&SU
multipoles) from the {\it full}
experimental total photoabsorption
cross section.  Of course, this multi-pion piece is presumably
associated with both $\Delta$P={\sc yes} and $\Delta$P={\sc no}
multipoles, and
these two components contribute with opposite signs
to the dispersion integral.  We can obtain an upper or lower
bound to the contribution of the multi-pion final states by
assuming that the multi-pion photoabsorption is {\it completely} $\Delta$P={\sc
yes}
or $\Delta$P={\sc no}, respectively.   We have applied this procedure using
two different
compilations of the experimental total photoabsorption cross-section,
one due to Damashek and Gilman\cite{Da70} and the other due to Armstrong
\cite{Ar72}.
These give similar results for the dispersion integral.  The integrand is
shown in Fig.~\ref{fig:schan}.  In this way we find for the s-channel
multi-pion contribution
\begin{eqnarray}
\bar{\alpha}_E-\bar{\beta}_M&=&\begin{array}{ll}
                                 \pm 3.0 &  \mbox{[s-channel multi-pion]},
\end{array}
\label{eq:multipi}
\end{eqnarray}
where the positive sign applies if the multi-pion photoabsorption is purely
$\Delta$P={\sc yes} (such as would obtain if the multi-pion part
were principally $\pi\Delta$ production in a relative S-state) and the
negative sign applies if it is purely $\Delta$P={\sc no} (such as would obtain
if the multi-pion part
were principally non-resonant $\pi\pi N$ with everything in a
relative S-state).  The value quoted is for the proton since only in this
case is there available a full tabulation of the total photoabsorption cross
section.
However, it is reasonable to assume that the neutron contribution would be
similar.

In the absence of additional experimental information on the multipole
content of the multi-pion final states, the only way to improve on the above
bounds is in the context of a model.  One such model is that
due to L'vov \cite{Lv79,Lv81}, wherein $\pi\pi N$ production is
approximated by the pion pole contribution to the process $\gamma
N\rightarrow\pi\Delta\rightarrow\pi\pi N$, and the amplitudes for all but
the relative $\pi\Delta$ S-wave are calculated in the Born approximation.
The S-wave component is adjusted so that the total $\pi\pi N$ cross section
evaluated in this manner agrees
with experiment \cite{Lv81}.  Using the computer code supplied by L'vov, we
have calculated these amplitudes and used them as input to the s-channel
integral, the integrand of which is shown in Fig.~\ref{fig:schan1}.  We find
\begin{eqnarray}
\bar{\alpha}_E-\bar{\beta}_M&=&\left\{\begin{array}{ll}
                                  +1.66 & \mbox{[s-channel multi-pion-
$\Delta$P={\sc yes}]} \\
                                  -1.10 &  \mbox{[s-channel multi-pion-
$\Delta$P={\sc no}]},
                                 \end{array}\right.
\label{eq:lvov}
\end{eqnarray}
with identical values for the neutron and proton, since such an approach
yields a strictly isoscalar amplitude.
We first note that the sum of the magnitudes of the two contributions (2.76) is
slightly less than the value of 3.0 obtained above.  Presumably this is due to
the neglect of final states with three or more pions in the model.  We further
note that there is considerable cancellation between the $\Delta$P={\sc yes}
and $\Delta$P={\sc no} components in the model calculation, so that the net
contribution of the multi-pion final states is quite small (0.56).  We return
to
this point below when we compare with experiment.

For the t-channel integral, we require the amplitudes for both
$\pi\pi\rightarrow N\bar{N}$ and $\gamma\gamma\rightarrow \pi\pi$.
The former can be reliably obtained by extrapolation from the
cross-channel
$\pi N\rightarrow \pi N$ process as done by
Bohannon and Signell \cite{Bo74}, from which we take the amplitudes
$f_+^{0;2}(t)$ for use in Eq.~(\ref{eq:tchan1}).  Previous calculations of the
backward sum rule have utilized these same forms.  The amplitudes for
$\gamma\gamma\rightarrow\pi\pi$ have traditionally been considered less
reliable, especially for the S-wave part.  However, in recent years there has
been renewed interest in this reaction from both the experimental and
theoretical side.  In particular, new calculations lead to
cross sections that are in excellent agreement with experimental data for
both the $\pi^+\pi^-$ and $\pi^0\pi^0$ channels \cite{Do93}.  These
calculations utilize dispersion relations with subtraction constants fixed by
low energy theorems, taking into account the $\pi\pi$ scattering phase shifts
as well as the effects of $\pi$, $\rho$,
$\omega$, and A$_1$ exchange \cite{Do93}.  At very low energy, the amplitude is
dominated by the pion Born and polarizability terms,
\begin{eqnarray}
F_0^0(t)&=&{16\pi\alpha m_\pi^2\over \sqrt{t(t-4m_\pi^2})}\ln {t+
\sqrt{t(t-4m_\pi^2)}\over t-\sqrt{t(t-4m_\pi^2)}}+4\pi m_\pi
t\bar{\alpha}_E^\pi
+\ldots
\label{eq:pipiggs}
\end{eqnarray}
for S-wave and
\begin{eqnarray}
F_0^2(t)&=&40\pi\alpha\left[-{6m_\pi^2\over t-
4m_\pi^2}+2\left(1+{6m_\pi^2\over
t-4m_\pi^2}\right){m_\pi^2\over \sqrt{t(t-4m_\pi^2)}}\ln{t+\sqrt{t(t-
4m_\pi^2)}
\over t-\sqrt{t(t-4m_\pi^2)}}\right]\nonumber\\
\quad
\label{eq:pipiggd}
\end{eqnarray}
for D-wave.  Here
$\bar{\alpha}_E^\pi$ is the electric polarizability of the charged pion, for
which
there is a precise prediction of  $(2.8\pm 0.3)$
based on chiral symmetry, with parameters fixed from radiative pion decay
\cite{Te72}.
The fact that neither the chiral prediction nor
the $\pi\pi\rightarrow \gamma\gamma$ results are expected to be
accurate for
energies $E\ge$ 600 MeV is not a significant
problem as the factor t$^{-2}$ in Eq.~(\ref{eq:tchan1}) guarantees rapid
convergence
of
the dispersion integral (see Fig.~\ref{fig:tchans}).  The contributions to the
proton and neutron
integrals are
identical as only the isoscalar $N\bar{N}$ channel is allowed by G-parity
invariance.

In our numerical t-channel calculation, we include
only S- and D-wave components.  The integrand for the
S-wave piece is given in Fig.~\ref{fig:tchans}, in which three
different
curves are shown, corresponding to three different representations of the
$\pi\pi\rightarrow \gamma\gamma$ amplitude:  the Born approximation
(the polarizability-independent term in Eq.~(\ref{eq:pipiggs})), Born plus
pion
polarizability (Eq.~(\ref{eq:pipiggs})), and the full dispersively
calculated amplitude.  The chiral
prediction
for $\bar{\alpha}_E^\pi$ is used.  We see that
the
full amplitude looks significantly different from the other two, mainly
because of a zero in F$_0^0$(t) near 400 MeV which arises due to the Omnes
function for I=0 $\pi\pi $ scattering \cite{Do93}.  The numerical results
\begin{eqnarray}
\bar{\alpha}_E-\bar{\beta}_M&=&\left\{\begin{array}{ll}
+16.1&\mbox{[t-channel S-wave, Born]}\\
+19.1&\mbox{[t-channel S-wave, Born + $\bar{\alpha_E^\pi}$]}\\
+10.3&\mbox{[t-channel S-wave, full]}\end{array}\right.
\label{eq:ts}
\end{eqnarray}
are quite sensitive to the location of this zero, which explains much of the
uncertainty in the previous calculations of this contribution.  Nevertheless,
the excellent agreement between the full amplitude and the recent cross
section data \cite{Do93} gives us confidence in our result.

For the D-wave piece we use the Born approximation for
$\pi\pi\rightarrow \gamma\gamma$ (Eq.~(\ref{eq:pipiggd})).  The
integrand
is shown in Fig.~\ref{fig:tchand}; the integral
\begin{eqnarray}
\bar{\alpha}_E-\bar{\beta}_M&=&\begin{array}{ll}
                                 -1.7&  \mbox{[t-channel D-wave]},
\end{array}
\label{eq:td}
\end{eqnarray}
is significantly smaller than its S-wave counterpart, thereby providing some
justification for the neglect of higher partial waves.  We note that the
magnitude of
our D-wave contribution is nearly a factor of four smaller than that
given by previous authors \cite{Be74,Bu79}.  We do not understand the origin
of this discrepancy.

Putting everything together we arrive then at our final results
\begin{eqnarray}
\bar{\alpha}_E^p-\bar{\beta}_M^p&=&\left\{\begin{array}{ll}
                                   5.6 & \mbox{[upper bound]} \\
                                   3.2 &  [\pi\Delta \mbox{ model}]\\
                                   -0.4 & \mbox{[lower bound]} %\\
%                                   8.4\pm~2.1 & \mbox{[experiment]}
                                   \end{array}\right.\nonumber\\
\bar{\alpha}_E^n-\bar{\beta}_M^n&=&\left\{\begin{array}{ll}
                                   6.3 & \mbox{[upper bound]} \\
                                   3.9 & [\pi\Delta \mbox{ model]}\\
                                   0.3 & \mbox{[lower bound]}% \\
%                                   9.4\pm~5.0 & \mbox{[experiment]}
                                   \end{array}\right.
\label{eq:results}
\end{eqnarray}
These numbers are to be compared with the experimental values:
\begin{eqnarray}
\bar{\alpha}_E-\bar{\beta}_M&=&\left\{\begin{array}{ll}
8.4\pm 2.1&\mbox{[experiment p]}\\
9.4\pm 5.0&\mbox{[experiment n]}
\end{array}\right.
\label{eq:diffexpt}
\end{eqnarray}
Taking into account the errors\footnote{To obtain
the error on $\bar{\alpha}_E-\bar{\beta}_M$, we first combine in
quadrature the
errors on $\bar{\alpha}_E$ in Eq.~(\ref{eq:polarizabilities}), then double the
result,
since the errors on $\bar{\alpha}_E$ and $\bar{\beta}_M$ are
anticorrelated \cite{Na94}.} on the experimental results, there is good overall
consistency with the backward sum rule, provided the actual contribution of
the
s-channel multi-pion contribution is somewhere between the upper bound
and the
$\pi\Delta$ model prediction.  However, additional work would be very
helpful in
extending these
findings.  In particular
a multipole analysis of the $\gamma N\rightarrow \pi\pi N$ process, such
as is
presently planned at Argonne \cite{Le92},
would help to clarify the full s-channel dispersive analysis.

We now return to the point that originally motivated this work, the size of
$\bar{\alpha}_E^n$ relative to $\bar{\alpha}_E^p$.  Combining the Baldin
and
backward sum rules, we obtain
\begin{equation}
\bar{\alpha}_E\,=\,{1\over 2}\left[(\bar{\alpha}_E+\bar{\beta}_M)+
(\bar{\alpha}_E-\bar{\beta}_M)\right],
\end{equation}
and then take the neutron-proton difference of these quantities.  The
dispersion
prediction should be particularly accurate for this difference because the
principal
uncertainties in our calculation are in the s-channel multi-pion contribution,
which is
approximately isoscalar, and the t-channel contribution, which is rigorously
isoscalar.
Therefore, those uncertainties are largely removed when we take the
neutron-proton
difference.  We find the following:
\begin{eqnarray}
\bar{\alpha}_E^n-\bar{\alpha}_E^p&=&\left\{\begin{array}{lll}
1.2&\mbox{[dispersion relations]}\\
1.3\pm 1.9&\mbox{[experiment]}\\
-3.6&\mbox{[valence quark model]}
\end{array}\right.
\end{eqnarray}
We see that the dispersion theory does remarkably well in quantitatively
accounting
for the relative sizes of the electric polarizability for the neutron and
proton.
It
appears
that both the chiral perturbative, Eq.~(\ref{eq:chiral}), and dispersive
calculations
are quite consistent
with the experimental findings, while
the simple constituent quark model, Eq.~(\ref{eq:quark}), is strongly at
variance.  We conclude that
taking the pion cloud components of the nucleon into account is essential in
order to understand the recent polarizability results.

\section{ Connecting Pion and Nucleon Polarizabilities}

Since the electric  and the diamagnetic polarizabilities are
dominated by the pion cloud, it is reasonable to ask whether the intrinsic
polarizability of the pion itself contributes to that of the nucleon.
Intuitively
such a connection is expected since the presence of an external
electromagnetic field can not only polarize the pion cloud relative to the
quark core but can also, to the extent that the pion is polarizable, polarize
the
pions themselves.  The backward dispersion relation enables us to derive a
model-independent relation between the nucleon and pion polarizabilities,
which can be compared to the predictions of various models.  In this section
we address this issue.

Cohen and Broniowski have derived a quantitative relation between the
nucleon and pion polarizabilities in the context of a hedgehog model of the
nucleon \cite{Br93}.  They focus on
the
$L_9,L_{10}$ component---the piece responsible for giving the pion
electromagnetic structure---of the effective action describing the interaction
of Goldstone bosons, as written down by Gasser and Leutwyler\cite{Ga84}
\begin{eqnarray}
{\cal L}_{\rm eff}=\ldots -iL_9{\rm Tr}\left[F^L_{\mu\nu}D^\mu UD^\nu
U^\dagger
+F^R_{\mu\nu}D^\mu U^\dagger D^\nu U\right]\nonumber\\
\quad +L_{10}{\rm
Tr}\left[F^L_{\mu\nu}UF^{R,\mu\nu}U^\dagger\right],
\end{eqnarray}
where $F^{L,R}_{\mu\nu}$ are the left,right chiral field strength tensors,
which
in the electromagnetic case take the form $F^{L,R}_{\mu\nu}={e\over 2}
\tau_3F_{\mu\nu}$.  In the linear sigma model, $U$ describes the chiral
field, $U={1\over F_\pi}(\sigma+i{\bf \tau}\cdot{\bf \pi})$.  At tree level
the charged pion polarizability can be completely described in terms of
$L_9,L_{10}$:
\begin{eqnarray}
L_9&=&{F_\pi^2<r_E^2>\over 12}\nonumber\\
L_{10}&=&{m_\pi F_\pi^24\pi\bar{\alpha}_E^\pi\over 4e^2}-L_9\equiv{ m_\pi
F_\pi^2
4\pi\alpha_E^\pi \over 4e^2}\nonumber\\
& & {\rm with}\quad \bar{\alpha}_E^\pi =-\bar{\beta}_M^\pi
\end{eqnarray}
In a mean field approach, treating the meson operators as classical fields
and taking ${\bf E},{\bf B}$ to be constants, one finds
\begin{equation}
\int d^3x{\cal L}\sim \int d^3x {4e^2\over F_\pi^2}(L_9+L_{10})({\bf
E}^2
-{\bf B}^2)({\bf c}\times\pi_h)^2
\end{equation}
where ${\bf \pi}_h$ is the
hedgehog pion field and ${\bf c}$ is defined if ref. 32.
The spatial integral can be related to the
fraction of the total moment of inertia carried by the pion degree of
freedom and yields the estimate\cite{Br93}
\begin{equation}
\delta\bar{\alpha}_E^N= -\delta\bar{\beta}_M^N\approx
0.5\bar{\alpha}_E^\pi,
\label{eq:hedgehog}
\end{equation}
where $\delta\bar{\alpha}_E^N$ refers to that part of the nucleon
polarizability that is due to the intrinsic polarizability of the pion.

It is possible to understand this result in an alternative fashion,
using Feynman
diagrams.  Thus the effective charged pion electromagnetic interaction
due to its polarizability can be written in the local form\cite{Ho90a}
\begin{equation}
{\cal L}^\pi_{\rm eff}={1\over
4}F_{\mu\nu}F^{\mu\nu}4\pi\bar{\alpha}_E^\pi 2m_\pi
\pi^+\pi^-.
\end{equation}
Insertion into the diagram shown in Fig.~\ref{fig:diagram} then yields
\begin{eqnarray}
{\cal L}^N_{\rm eff}&=&{1\over
4}F_{\mu\nu}F^{\mu\nu}4\pi\bar{\alpha}_E^\pi 2m_\pi
(\sqrt{2}g)^2\int{d^4k\over (2\pi )^4}{1\over (k^2-m_\pi^2)^2}\bar{\psi}
\gamma_5
{1\over \gamma^\mu (p-k)_\mu -m_N}\gamma_5 \psi \nonumber\\
&=&F_{\mu\nu}F^{\mu\nu}\bar{\psi}\psi\times
4\pi\bar{\alpha}_E^\pi({g\over 4\pi})^2 r_\pi I(r_\pi^2 )
\end{eqnarray}
where $r_\pi=m_\pi /m_N$ and
\begin{equation}
I(x)=\int_0^1 dy {y(1-y)\over y^2+x(1-y)}.
\end{equation}
We identify then the contribution to the nucleon polarizability due to
the analogous pion polarizability as
\begin{equation}
\delta\bar{\alpha}_E^N=-\delta\bar{\beta}_M^N=4({g\over 4\pi})^2r_\pi
I(r_\pi^2)
\bar{\alpha}_E^\pi
=0.8\bar{\alpha}_E^\pi,
\label{eq:feynmann}
\end{equation}
which is somewhat larger than the hedgehog number.  However, it must be
emphasized
that this is a simple one loop calculation and must therefore be considered
to be only a crude estimate.

Finally, we derive a basically model-independent result based on the backward
dispersion
relation.  The connection comes via the t-channel integral,
Eq.~(\ref{eq:tchan1}), and the low-energy form of the S-wave part of the
$\gamma\gamma\rightarrow\pi\pi$ amplitude, Eq.~(\ref{eq:pipiggs}),
from which one easily derives
\begin{equation}
\delta\left(\bar{\alpha}_E^N-\bar{\beta}_M^N\right)=\bar{\alpha}_E^\pi
{4m_\pi\over \pi}
\int_{4m_\pi^2}^\infty {dt\sqrt{t-4m_\pi^2}\over t^{3\over 2}(4m_N^2-
t)}|f_+^{0}(t)|,
\end{equation}
Numerical evaluation of this integral then gives\footnote{This procedure has
been looked
at previously by V.M. Budnev and V.A. Karnakov \cite{Bu79}.  However,
there appear to exist a number of serious
dimensional errors in their paper ({\it cf.} Eqs. 4
and 9) so that the numerical values given therein must be questioned.}
\begin{equation}
\delta\left(\bar{\alpha}_E^N-
\bar{\beta}_M^N\right)=1.01\bar{\alpha}_E^\pi
\label{eq:need}
\end{equation}
Since from our previous discussion the pion contribution to the nucleon
electric/magnetic polarizabilities is equal and opposite, we
can rewrite Eq.~(\ref{eq:need})as
\begin{equation}
\delta\bar{\alpha}_E^N=-\delta\bar{\beta}_M^N=0.5\bar{\alpha_E^\pi}
\label{eq:disp}
\end{equation}
which is a rigorous result and in satisfactory agreement with the estimates
provided above via hedgehog and Feynman diagram arguments.

The size of this contribution to the nucleon polarizability depends
upon the size of the charged pion polarizability, whose value is
still experimentally uncertain.  Although chiral symmetry makes a rather firm
theoretical prediction \cite{Te72}
\begin{equation}
\bar{\alpha}_E^\pi\,=\,-\bar{\beta}_M^\pi\,=
{4\alpha(L_9+L_{10}\over m_\pi F_\pi^2}=\,2.8\,\mbox{ [chiral prediction],}
\end{equation}
the experimental situation is yet unclear with three different results
being provided by three very different techniques:
\begin{eqnarray}
\bar{\alpha}_E^\pi &=&\left\{\begin{array}{lll}
2.2\pm 1.1&\mbox{[$\gamma\gamma\rightarrow\pi\pi$]}&\cite{Ba92}\\
6.8\pm 1.4&\mbox{[radiative pion scattering]}&\cite{An83}\\
20\pm 12&\mbox{[radiative pion photoproduction]}&\cite{Ai86}
\end{array}\right.
\end{eqnarray}

Comparison with the recently measured nucleon values
Eq.~(\ref{eq:polarizabilities})
indicates that the pion contribution to the nucleon polarizability is
relatively
modest if the chiral prediction or the $\gamma\gamma\rightarrow\pi\pi $
result is correct, but is rather significant if the radiative pion
scattering value were to be correct.  Note that since the t-channel
dispersive piece is isoscalar, its contribution to neutron and proton
values is identical.  Clearly it is important to remeasure
the pion polarizability in order to resolve the origin of these
discrepant values, and such efforts are planned at
Brookhaven, Fermilab, and Da$\Phi$ne.

\section{Conclusions}
Recent experimental measurements of the nucleon electromagnetic
polarizabilities are shown to be inconsistent with expectations based
on a simple constituent quark model picture of the nucleon---mesonic
contributions must be included in order to understand
these findings.  Dispersion relations offer a rigorous approach to
this problem, but depend sensitively upon the correctness of the
s- and t-channel integrands.  Considerable recent progress has been
made in this regard.  In the case of the t-channel, successful dispersive/
chiral perturbative analyses of the $\gamma\gamma\rightarrow\pi\pi$
reaction
have enabled a reasonably reliable estimate of this contribution, while in
the case of the s-channel a multipole analysis of the $N\pi$ intermediate
state enables a believable calculation of this piece.  Further
progress awaits a similar multipole decomposition of the (smaller)
multi-pion component as well as an improvement on the precision of the
neutron polarizability measurements.  However, overall agreement between
the experimental
numbers and the dispersive predictions must be judged to be quite
satisfactory.  Finally, we have used the t-channel part of the dispersion
relation to do a precise calculation of the contribution of the pion
polarizability to that of the nucleon.

\medskip

{\bf Acknowledgement:}  It is a pleasure the acknowledge very fruitful
conversations with A. I.  L'vov as well the use of his computer code for the
calculation of the model 2-pion photoproduction multipole amplitudes.  BRH
thanks the Institute for Nuclear Theory and AMN thanks the Nuclear
Physics Laboratory at the University of Washington for their hospitality
during the initial stages of this work.

%
%	REFERENCES
%

%
%	FIGURE CAPTIONS
%
\newpage
\begin{center}
Figure Captions
\end{center}
Figure 1:  Integrand of the s-channel contribution for the proton.  The
solid/dashed curves are the integrans for the single-pion parity
changing/non-changing multipoles, respectively.  The dotted vurve is the
integrand obtained by subtracting the single pion from the total
photoproduction cross section, and the integral of that curve gives a
rigorous bound on the multi-pion contribution.\\

Figure 2:  The integrand of the s-channel multipion contributions for the
proton, as given by the $\pi\Delta$ model of L'vov.  The solid/dashed
curves are the integrands for the two-pion parity changing/non-changing
multipoles, respectively.\\

Figure 3:  The integrand for the t-channel S-wave contribution.  The
solid, dotted, and dashed curves correspond to the Born, Born+pion
polarizability and full amplitudes, respectively, for the
$\gamma\gamma\rightarrow\pi\pi$ process.\\

Figure 4:  The integrand for the t-channel D-wave contribution.\\

Figure 5:  Pion-exchange diagram contributing to the nucleon polarizability.\\


\begin{thebibliography}{99}
\bibitem{Fe91} F.J. Federspiel et al., Phys. Rev. Lett. {\bf 67}, 1511 (1991).
\bibitem{Zi92} A. Zieger et al., Phys. Lett. {\bf B278}, 34 (1992).
\bibitem{Ha93} E. Hallin, et al., Phys. Rev. C.{\bf 48}, 1497 (1993).
\bibitem{Sc91} J. Schmiedmyer et al., Phys. Rev. Lett. {\bf 66}, 1015 (1991).
\bibitem{Na94} A. M. Nathan, A. I. L'vov, V. A. Petrun'kin, to be published.
\bibitem{Fr75} J. L. Friar, Ann. Phys. (N.Y.) {\bf 95}, 170 (1975).
\bibitem{Mu93}N. C. Mukhopadhyay, et al., Phys. Rev. D {\bf 47}, 7 (1993).
\bibitem{Ho92}B. R. Holstein, Comm. Nucl. Part. Phys. {\bf 20}, 201 (1992).
\bibitem{Lv93a}A. I. L'vov in {\it Baryons `92}, ed. M. Gai (World Scientific,
Singapore, 1993).
\bibitem{We85} R. Weiner and W. Weise, Phys. Lett. {\bf 159B}, 85 (1985).
\bibitem{We88} T.E.O. Ericson and W. Weise, {\it Pions and Nuclei} (Clarendon
Press, New York, 1988).
\bibitem{Be91} V. Bernard et. al., Phys. Rev. Lett. {\bf 66}, 1515 (1991).

\bibitem{Lv93} A. I. L'vov, Phys. Lett. {\bf 304B}, 29 (1993).

\bibitem{Ho90} See, e.g., B.R. Holstein in {\it Parity Violation in Electron
Scattering}, ed. E. J. Beise and R. D. McKeown (World Scientific, Singapore,
1990).

\bibitem{Pe81} V. A. Petrun'kin, Sov. J. Part. Nucl. {\bf 12}, 278 (1981).

\bibitem{Ba60} A. M. Baldin, Nucl. Phys. {\bf 18}, 318 (1960).

\bibitem{Da70} M. Damashek and F.J. Gilman, Phys. Rev. D {\bf 1}, 1319
(1970).

\bibitem{Lv79}A. I. L'vov, V. A. Petrun'kin and J. A. Startsev, Sov. J. Nucl.
Phys. {\bf 28}, 651 (1979).

\bibitem{Be74}J. Bernabeu, T. E. O. Ericson and C. Ferro Fontan, Phys.
Lett. {\bf 49B}, 381 (1974); J. Bernabeu and B. Tarrach, Phys. Lett. {\bf
69B}, 484 (1977).
\bibitem{Ch57} G. F. Chew, M. L. Goldberger, F. E. Low and Y. Nambu,
Phys. Rev. {\bf 106}, 1345 (1957).
\bibitem{Go60} M. Gourdin and A. Martin, Nuovo Cimento {\bf 17}, 224
(1960).
\bibitem{Fr60} W.R. Frazer and J.R. Fulco, Phys. Rev. {\bf 117}, 1603 (1960).
\bibitem{Gu78} I. Guia\c{s}u and E. E. Radescu, Phys. Rev. D {\bf 18}, 1728
(1978).
\bibitem{Bu79} V. M. Budnev and V. A. Karnakov, Sov. J. Nucl. Phys. {\bf
30}, 228 (1979).
\bibitem{Ar90} R. A. Arndt, et al., Phys. Rev. C {\bf 42}, 1853 (1990).
\bibitem{Ar72} T. A. Armstrong, et al., Phys. Rev. D {\bf 5}, 1640 (1972).
\bibitem{Lv81} A. I. L'vov et. al., Sov. J. Nucl. Phys. {\bf 29}, 651 (1981).
\bibitem{Bo74} G.E. Bohannon and P. Signell, Phys. Rev. D {\bf 10}, 815
(1974).
\bibitem{Do93} J. F. Donoghue and B. R. Holstein, Phys. Rev. D {\bf 48}, 137
(1993).
\bibitem{Te72} M.V. Terent'ev, Sov. J. Nucl. Phys. {\bf 16}, 162 (1972);
J.F. Donoghue and B.R. Holstein, Phys. Rev. D {\bf 40}, 2378 (1989).
\bibitem{Le92} T. S. H. Lee, private communication.
\bibitem{Br93} W. Broniowski and T.S. Cohen, Phys. Rev. D {\bf 47}, 299
(1993).
\bibitem{Ga84} J. Gasser and H. Leutwyler, Ann. Phys. (NY), {\bf 158}, 142
(1984);
Nucl. Phys. {\ bf B250}, 465 (1985).
\bibitem{Ho90a} B.R. Holstein, Comm. Nucl. Part. Phys. {\bf 19}, 221 (1990).
\bibitem{Ba92} D. Babusci et al., Phys. Lett. {\bf B277}, 158 (1992).
\bibitem{An83} Yu. M. Antipov et al., Phys. Lett {\bf B121}, 445 (1983);
Z. Phys. {\bf C26}, 495 (1985).
\bibitem{Ai86} T.A. Aibergenov et al., Czech. J. Phys. {\bf B36}, 948 (1986).
\end{thebibliography}
\end{document}